\documentclass[prd,preprint,a4paper,superscriptaddress,nofootinbib,11pt]{revtex4}
\usepackage{amsmath,amsfonts,amssymb}
\usepackage{txfonts}
\usepackage[T1]{fontenc}
\usepackage[safe]{textcomp}
\usepackage{graphicx}
\usepackage{physics}
\usepackage[left=2cm,right=2cm,top=2cm,bottom=2cm]{geometry}
\usepackage{cancel}
\usepackage[toc,page]{appendix}
\usepackage{epstopdf, epsfig}

\newcommand{\be}{\begin{equation}}
\newcommand{\ee}{\end{equation}}
\newcommand{\bea}{\begin{eqnarray}}
\newcommand{\eea}{\end{eqnarray}}
\newcommand{\ba}{\begin{aligned}}
\newcommand{\ea}{\end{aligned}}
\newcommand{\ks}{$\kappa$-spacetime}
\newcommand{\kdst}{$\kappa$-deformed spacetime}

\newcommand{\hu}[2]{\hat{#1}^{#2}}
\newcommand{\hd}[2]{\hat{#1}_{#2}}
\renewcommand{\comm}[2]{[ #1, #2 ]}

\newcommand{\kd}{$\kappa$-deformed}

\newcommand{\sch}{Schwarzschild}

\begin{document}
\author{E. Harikumar\footnote{harisp@uohyd.ernet.in}, Zuhair N. S.\footnote{zuhairns@gmail.com}}
\affiliation{School of Physics, University of Hyderabad, \\ Central University P.O., Gachibowli, Telangana, India - 500046}
\title{Hawking radiation in $\kappa$-spacetime}

\begin{abstract}
In this paper, we analyze the Hawking radiation of a \kd \, \sch \, black hole and obtain the deformed Hawking temperature. For this, we first derive deformed metric for the \ks , which in the generic case, is not a symmetric tensor and also has a momentum dependence. We show that the \sch \, metric obtained in the \kdst \, has a dependence on energy. We use the fact that the deformed metric is conformally flat in the 1+1 dimensions, to solve the \kd \, Klein-Gordon equation in the background of the \sch \, metric. The method of Boguliobov coefficients is then used to calculate the thermal spectrum of \kd -\sch \, black hole and show that the Hawking temperature is modified by the non-commutativity of the \ks .
\end{abstract}
\maketitle
\section{Introduction}
A proper understanding of spacetime is indispensable for formulating a quantum theory of gravity. There are convincing arguments that usual notion of spacetime breaks down as we move up in energy scale \cite{dop, raam, amelino, nima, horowitz}. In fact, many interesting ideas about the possible structure of spacetime at the Planck regime have been brought up by several authors (see \cite{GSW,rov, perez, cdt, connes} and referenes therein). One of the possible candidates to study the spacetime structure at the Planck scale is the non-commutative spacetime approach. In addition, the need for modifying special theory of relativity to incorporate an absolute length scale leads to a deformed relativity principle which also leads to a non-commutative spacetime and the corresponding spacetime called \kdst \, is studied in \cite{luk, kowalski}.

One of the best place to investigate the effect of deformation of spacetime would be near a black hole. In the vicinity of a black hole, gravity is so strong that any deformation of the spacetime structure is expected to be tractable. Now, for an inertial observer who is at a finite distance from the event horizon, the black hole appears to radiate particle with a thermal spectrum. The presence of this thermal radiation (known as Hawking radiation) from black hole is a key signature of it's quantum nature\cite{hawking}. The thermal radiation of black hole is such that the temperature of thermal spectrum would be inversely proportional to the mass of the black hole. Now, as the black hole radiates it looses energy and its mass reduces. As it looses mass, its temperature rises and this process continues. This led to the information loss problem and poses one of the major crisis in the understanding of theoretical physics \cite{wald}. In order to precisely understand the situation we need to take into account of the full quantum gravity effects. In other words, Hawking radiation reveals the thermodynamic behaviour of spacetime (see \cite{wald} and references therein). By studying the thermal distribution of the black hole we could look for any possible deviation of spacetime structure from the expected ones. 

Since non-commutativity is speculated by various theories (see \cite{string, loop} and references therein) as a plausible nature of spacetime at the Planck scale, it would be interesting to investigate the characteristics of a black hole in non-commutative spacetimes. Black holes and its thermodynamic behaviour is investigated in Moyal space and it is argued that a possible resolution for black hole information paradox can be achieved within the non-commutative framework. In fact, it is claimed that  the curvature singularity vanishes in non-commutative spacetime and black hole emits radiation till it becomes a zero temperature extremal black hole \cite{NCBH1}. Non-commutativity is viewed as smearing of spacetime points and non-commutative geometry near black holes is studied using smeared sources rather than point like sources \cite{NCBH2, NCBH3}. A detailed analysis of non-commutative black hole in Moyal space has been given in \cite{NCBH4}. Different aspects of black hole in \ks \, have been studied in \cite{ksg1, ksg2}. A phenomenon of very close relationship to Hawking radiation is Unruh effect. \cite{unruh, fulling, unruha, singleton}. Basically, Unruh effect states that, an accelerating observer measures a thermal radiation in an inertial vacuum. For some of the recent studies of Unruh effect in non-commutative spacetime, see \cite{unruh1, hariravi1, hariravi2, unruh2}. 
 
In standard calculations of Hawking radiation, dynamics of background geometry is not taken into account. An alternative approach to 
calculate Hawking radiation is presented in \cite{parwil} by viewing background as dynamical and explaining the process of Hawking 
radiation in terms of quantum tunnelling. The method presented in \cite{parwil} leads to a correction to the spectrum of black hole 
radiation and leads to a non-thermal spectrum. Extension of tunnelling approach to study the Hawking radiation in non-commutative 
spacetime leads to a corrected spectrum for Hawking radiation and hopes to cure the information problem \cite{ncbh}. By considering a 
Lorentz violating dispersion relation which results in a momentum dependent metric, it was shown that the black hole evaporation ends 
at a non-zero critical mass \cite{energymetric}. Energy dependent metric has appered in the context of deformed special relativity also
\cite{GP}.

Considering different approaches in non-commutative black holes, most of the studies were carried out in Moyal spacetime 
(\cite{NCBH1, NCBH2, NCBH3, NCBH4, ncbh} and references therein). In this paper, we analyze the Hawking radiation in a specific 
non-commutative spacetime called \ks . We use a mapping connecting the non-commutative coordinates to commutative coordinates 
and momenta. Exploiting this map, the problem of Hawking radiation in $\kappa$-spacetime is mapped to an equivalent problem in the 
commutative spacetime. This mapping between non-commutative coordinates and coordinates of commutative phase space is derived by 
demanding the consistency of the algebra of non-commuting variables. This mapping is used to obtain a commutative models equivalent 
to the given \kd \, model \cite{mel, hari1}. This approach was used to analyze Kepler system in \kdst \, and it was found that the 
deformed problem retains all the symmetries as the commutative Kepler problem\cite{zns,zns1,zns2}. Using this approach, change in the 
dimension of \ks \, with probe scale was analyzed in \cite{ anjana, anjana1, anjana2}.

In this paper, $\kappa$-deformed black hole is chosen as a platform to investigate the features arising from the kappa non-commutativity. We adopt a generic method to write a metric element in the $\kappa$-deformed spacetime. The method is then used to investigate the deformation of the metric of the \kd \,  Schwarzschild black hole spacetime. We then use a specific realization of \ks \, variables to study the effects of $\kappa$-deformation on the Hawking temperature. Now, we would like to highlight three key features of our result. First feature is that we have obtained Hawking effect using the idea that the particles in \ks \, obeys a twisted statistics.

 It was shown in \cite{twist, twist2} that the scalar field in the \ks \, do not satisfy $\phi (x) \phi (y) = \phi (y) \phi (y)$. This leads to modification of the commutation relations between creation and annihilation operators corresponding to the quantized scalar field in the \ks . We have used this information to obtain Hawking temperature in this paper.  We used the conformal flatness of the \kd \, \sch \, metric in 1+1 dimensions, to expand the scalar field in terms of its Fourier modes in two different coordinates bases, namely Kruskal-Szekeres coordinates and tortoise coordinates. The method of Boguliobov transformation is then used to relate the  creation and annihilation operators corresponding to the two coordinates. We then calculate the Boguliobov coefficients to obtain the expression for thermal radiation from the \kd -\sch \, black hole. Second point is that, we apply the procedure of Euclidean time formulation of the metric (see chapter 15 of \cite{euchawk}) to get the Hawking temperature in \kdst \, as elaborated in the appendix B. Given that, we had restricted ourselves to 1+1 dimensions in our derivation of Hawking radiation, the second method is to emphasize that the result obtained by us is independent of the dimension of spacetime under consideration. Third important characteristic is that, we find that the surface gravity of \kd -Schwarzschild black hole is modified by an exponential function of deformation parameter $a$. Hence, the Hawking temperature is also modified by the presence of an exponential factor which depends on the deformation parameter $a$. We have shown that the \kd \, metric do get energy and momentum dependent modifications in the \kdst . Further, all the results obtained fall back to familiar commutative results, when we let the deformation of spacetime to vanish. 

It is to be noted that there are different approaches in studying various aspects of gravity in the non-commutative framework. Deriving a non-commutative version of Einstein equation valid in non-commutative spacetime and analyzing thus obtained non-commutative Einstein equation is one approach.Another possibility is to start with a non-commutative counterpart of the commutative model of the system under consideration. Such non-commutative counter-parts of commutative systems are constructed primarily by imposing the restriction that in the limit of non-commutative parameter vanishing, we get back the well known commutative model. Non-commutative models constructed in this approach serve as a testing ground to study the possible effects of non-commutativity in the problem addressed. It is to cautioned that these non-commutative models may not be solution to the yet, unknown \kd \, Einstein equation. Thus, aim here is only to take a first step towards the study of modification of black hole physics in the $\kappa$-deformed spacetime. But this in itself is important as this provides insights in to the $\kappa$-deformation and its implications in black hole physics. Since, the current understanding of a \kd \,  Einstein's equation is incomplete, we choose the latter approach in this paper. 

There are two possible approaches for constructing and studying models in \kdst . Performing calculation using the functions of non-commutative coordinates is an option, while another is to work in terms of non-commutative functions expressed in terms of commutative variables. This second approach uses a mapping of non-commutative coordinates in terms of commutative coordinates, their derivatives and the non-commutative parameter. It is known that for \kdst , many different realisations of the non-commutative coordinates exist. It is also known that the choice of realisations is equivalent to the choice of ordering of non-commutative coordinates in defining functions in non-commutative spacetime and also related to the choice of star-product. Since there exist many realizations of $\kappa$-spacetime(see \cite{star} for more details), we have chosen a specific realization in this paper.

Organization of this paper is as follows. In section 2, we start with a particular choice of realization of non-commutative variables which map variables in \ks \, to variables in the commutative spacetime. This enables us to perform our computations using familiar commutative methods. We then utilize the general commutation relations for non-commutative coordinates to derive the deformation of the metric. We obtain the deformed metric for the \ks \, and find the corrections to the metric elements due to $\kappa$-deformation of spacetime. We show that the deformed metric depends on the energy and momentum, and when the deformation parameter is set to zero (i.e., $a \rightarrow 0$), all the additional contributions to the metric vanish. In the third section, we deal with the \kd -Schwarzschild metric. The deformed metric obtained after mapping to the commutative variables has the usual spherically symmetric form.  We investigate the temperature of the Hawking radiation for this metric using two different approaches. In the first case, we expand a massless scalar field in the background of the deformed metric in 1+1 dimensions. Here, it is to be emphasized that the conformal flatness of the deformed metric is exploited, so that all our calculations will be simplified to that of a flat metric. We now expand the Fourier modes of the field, with respect to two different choice of basis. The expectation value of number operator in one basis is then calculated in terms of vacuum state of other basis and the thermal distribution behaviour of the vacuum state is established. In the remaining section, we present a different route to this result. We use the path integral formulation with imaginary time method to identify the temperature with the surface gravity. Finally, in the last section, the effect of deformation factor is studied for various energy scales and an order of magnitude analysis is done for the deformation parameter, $a$. Our concluding remarks are also given here. In appendix A, we give details for deriving the realization of variable $\hat{y}$, used in our calculations, in terms of commuting variables. The variable $\hat{y}$ is introduced such that it commutes with the non-commuting coordinate $\hat{x}$ (see \cite{hari1, hari2}). Finally, in appendix B, we give a brief treatment of Hawking temperature in the Euclideanised metric. We first identify the point of singularity in the metric and then the line element is expressed around the point of singularity. The expression obtained after applying imaginary time formalism is compared with spherically symmetric deformed metric in the Euclidean space. The condition to avoid coordinate singularity is utilized to obtain the expression for inverse temperature.

 \section{Deformation of metric in \kdst}
In this section, we present a method to obtain the deformed metric in the \kdst . We demand that the variables in \kd \, phase space obey the commutation relation
\be 
[\hd{x}{\mu}, \hd{P}{\nu}] := i \hd{g}{\mu \nu},
\ee
and we identify the R.H.S. as the metric of the deformed spacetime. Then, we introduce an auxiliary variable $\hd{y}{\mu}$ such that it commutes with $\hd{x}{\mu}$ and in addition satisfies all the other properties as $\hd{x}{\mu}$. We then proceed to obtain the components of the deformed metric.

We choose a specific realization of \ks \, given by (for details see \cite{mel,hari1,kovac})
\be 
\hd{x}{\mu} = x_\alpha  \varphi_{\mu}^{\alpha}, \quad \hd{P}{\mu} = g_{\alpha \beta}(\hat{y})k^{\beta} {\varphi^{\alpha}}_{\mu}, \label{real1}
\ee
where $g_{\mu \nu}(\hat{y})$ possess the same functional form as the corresponding metric in the commutative space and $k_\beta$ is the momentum defined in the commutative spacetime, (for details see section 3.2 of \cite{hari1}). The metric $g_{\mu \nu}(\hat{y})$ is obtained by replacing the coordinates in the corresponding metric of commutative spacetime with the coordinates of $\hat{y}_\mu$. This $\hat{y}_{\mu}$ coordinates satisfying eqn. \eqref{ycomm}-\eqref{real2} are introduced to simplify the calculations (see \cite{hari2} for details).

The coordinates of the \kdst \, satisfy the commutation relations,
\be 
\comm{\hd{x}{0}}{\hd{x}{i}} = i a \hd{x}{i}, \quad \comm{\hd{x}{i}}{\hd{x}{j}} =0, \label{def1}
\ee
where $a$ is the deformation parameter. By demanding eqn. \eqref{real1} satisfy eqn. \eqref{def1}, we obtain
\be 
\varphi^{0}_0 =1, \, \, \varphi^{0}_i = 0 = \varphi^{i}_0, \, \, \varphi^{i}_j
= \delta^{i}_j e^{-ak^{0}}. \label{real} \ee

Note here that the realization $\varphi^{\mu}_\nu$ have no explicit $\vec{p}$ dependence unlike the realization used in \cite{hari2}.

The $\hd{y}{\mu}$ appearing in eqn. \eqref{real1} satisfy
\be 
\comm{\hd{y}{0}}{\hd{y}{i}} = i a \hd{y}{i}, \quad \comm{\hd{y}{i}}{\hd{y}{j}} =0,  \label{ycomm}
\ee
and the condition
\be 
\comm{\hd{y}{\mu}}{\hd{x}{\nu}} = 0, \label{cond2}
\ee
is used as a defining equation (see \cite{hari2}). 
We demand that $\hat{y}_\mu$ can be expressed in terms of commutative coordinates and momenta as
\be 
\hd{y}{\mu} = x_\alpha \phi^{\alpha}_\mu. \label{real2}
\ee
 Using eqn. \eqref{real2} in \eqref{cond2}, a straightforward computation yields us (see appendix)
  \bea 
  \hd{y}{0} &=& x_0 - a x_j k^{j}, \label{yhat2}\\
  \hd{y}{i} &=&  x_i. \label{yhat}
 \eea

 Once we have the expression of $\hd{y}{\mu}$, we are all set to find the deformation of metric in \ks. The metric can be derived from the expression \cite{hari1},
 \be 
 [\hd{x}{\mu}, \hd{P}{\nu}] \equiv i\hd{g}{\mu \nu}= i g_{\alpha \beta}(\hat{y}) \left( k^{\beta}\frac{\partial {\varphi^{\alpha}}_{\nu}}{\partial k^{\sigma}} {\varphi^{\sigma}}_{\mu} + {\varphi^{\alpha}}_{\mu} {\varphi^{\beta}}_{\nu} \right), \label{comm1}
 \ee 
 where we have used eqn.\eqref{real1}. The $\hd{g}{\mu \nu}$ so obtained, will give us the deformed metric. The deformed metric $g_{\mu \nu}(\hat{y})$ is obtained by replacing commuting coordinates with \kd \, coordinates in the metric of commutative spacetime and thus $g_{\alpha \beta}(\hat{y}) = g_{\beta\alpha}(\hat{y})$.
 
 Using the realizations given in eqns.\eqref{real1} and \eqref{real}, we calculate the R.H.S of eqn. \eqref{comm1} for each index combinations. Thus, we get
 \be
 [\hd{x}{0}, \hd{P}{0}] = i g_{00} (\hat{y}) \label{x0p0}
 \ee
 \be
[\hd{x}{0}, \hd{P}{i}] = i g_{i0} (\hat{y}) (1-a k^{0}) e^{-ak^{0}} - a g_{ik}(\hat{y}) k^{k} e^{-ak^{0}},
\ee
\be 
  [\hd{x}{i}, \hd{P}{0}] = i g_{i0} (\hat{y}) \exp(-ak^{0})
  \ee
  \be 
     [\hd{x}{i}, \hd{P}{j}] = i g_{ij} (\hat{y}) \exp(-2ak^{0}).  
 \ee

We now define the line element in \ks \, using the deformed metric as 
\be 
ds^{2} = \hd{g}{\mu \nu } d\hu{x}{\mu}d\hu{x}{\nu}, \label{lineelement}
\ee
where $\hat{g}_{\mu \nu}$ is deformed metric expressed in terms of commutative variables. The components of this deformed metric can be read-off from the eqns. \eqref{x0p0}-\eqref{lineelement} and are given by
\bea
\hd{g}{00 } &=&   g_{00} (\hat{y}),  \label{g00}\\
\hd{g}{0 i } &=& g_{i0} (\hat{y}) (1-ak^{0}) e^{-ak^{0}} - a g_{im}(\hat{y})  k^{m} e^{-ak^{0}}, \label{g0i}\\
\hd{g}{i0 } &=& g_{0i} (\hat{y}) \exp(-ak^{0}), \label{gi0}\\
\hd{g}{i j } &=& g_{ji} (\hat{y}) \exp(-2ak^{0}) \label{gij}.
\eea

Note that $\hat{g}_{\mu \nu}$ depends on the deformed four momenta through its dependence on $g_{\mu \nu}(\hat{y})$. A similar feature of the metric, with a dependence on energy and magnitude of spatial momentum vector, was seen in the context of study of evaporation of black hole \cite{energymetric}. It has been argued that such dependence in metric is natural consequence of presence of minimal length \cite{smolin}. An interesting feature of the metric in eqns.\eqref{g00}-\eqref{gij}, is that it has an explicit dependence on $k^{0}$ through the exponential terms appearing in the above equation and we will see later that this feature is significant for our result. In the limit $k^{0} \rightarrow \infty$ (keeping $a\neq 0$), we will only have time-like line elements as all of the metric elements except $\hat{g}_{00}$ vanishes. For the case, $a \neq 0 ,k^{0} \rightarrow 0$, the deformation of metric arise only through the realization of $\hat{y}_\mu$. When we let $k^{0}\rightarrow \infty, a \rightarrow 0$ (such that $ak^{0}=$constant), we see that $g_{00}(\hat{y})$ will reduce to the usual commutative metric and the deformation arise solely from terms with explicit $k^{0}$ dependence. It is interesting also to see that we recover the commutative metric in the limit $a \rightarrow 0$, independent of the value of $k^{0}$. 
A quick inspection of the metric elements reveal that the resulting non-commutative metric is asymmetric (see eqn.\eqref{g0i}-\eqref{gij}). This asymmetry is clearly a result of non-commutativity and as we let the deformation parameter to zero, we obtain a metric which is symmetric under the exchange of indices, as expected in the commutative spacetime. Furthermore, the metric element $g_{\mu \nu}(\hat{y})$ has the same functional form of  the metric element of the corresponding commutative spacetime but now expressed in terms of non-commutative variable. With our choice of realization (see eqns. \eqref{yhat2}, \eqref{yhat}), we find that there are two kinds deformations for the metric elements. First kind of deformation arise from the exponential factor arising in eqns. \eqref{g00} - \eqref{gij}. The second kind of deformation arise from the realization of $\hat{y}_\mu $ in terms of commuting variables. For our present realization, the second kind of deformation change only those metric elements that have explicit time dependence (see eqn. \eqref{yhat2}).

The explicit form of line element is thus
\be
\ba  
ds^{2} &= g_{00}(\hat{y}) dx^{0} dx^{0} \\ & +
  \left( g_{0i} (\hat{y}) (1-ak^{0}) e^{-ak^{0}} - a g_{ik}(\hat{y}) k^{k} e^{-ap^{0}} \right)  dx^{0} \exp(-ak^{0}) dx^{i} \\ & + g_{i0} (\hat{y}) \exp(-ak^{0}) \exp(-ak^{0}) dx^{i} dx^{0} \\ & + g_{ij} (\hat{y}) \exp(-4ak^{0}) dx^{i} dx^{j}.
\ea
\ee
We immediately recognise that the metric in \kdst \, is an asymmetric metric, with asymmetry arising from the cross term in space and time indices.  We have to keep in mind that the factors, $g_{\mu \nu}(\hat{y_i})$ has same form as the metric elements in usual commutative spacetime but now expressed in terms of coordinate $\hat{y_i}$. In our case, the choice of realization leads to $\hat{y}_i=x_i$ and this is to be contrasted with the situation studied in \cite{hari2}, where the realization of $\hat{y_i}$ has a dependence on deformation parameter.
\section{Schwarzschild metric in \ks \, and deformed Hawking temperature}

In this section, we derive Schwarzschild metric in \ks \, and using this we obtain modified Hawking temperature. We use the metric components obtained in the last section to write down the line element in \ks .  We first start with the Scwarzschild metric in the Minkowski spacetime,
\be 
ds^{2} =   -\left( 1-\frac{2GM}{r}\right) dt^{2} + \frac{1}{1-\frac{2GM}{r}} dr^{2}+ r^{2} d\Omega^{2}. 
\ee
It should be cautioned that above metric components is written in polar coordinates whereas the metric components we obtained in \eqref{g00}-\eqref{gij} are in Cartesian coordinates. Eqn.\eqref{yhat} tells us that, $g_{\mu \nu}(\hat{y_i})=g_{\mu \nu}(x_i)$. In other words, metric components, $g_{\mu \nu}(\hat{y_i})$, which depend only the non-commutative spatial coordinates $\hat{y}_i$, is the same as the metric element in commutative spacetime. Thus the metric that is of relevance for us would be of the form,
\be 
ds^{2} = g_{00}(\hat{y}) dx^{0} dx^{0} + g_{ij} (\hat{y}) \exp(-4ak^{0}) dx^{i} dx^{j}.
\ee
Note that we have assumed $c=1$ in the above equation. In order to write the \kd -\sch metric, first we have to transform the metric from Cartesian to polar coordinates. So the resulting metric in polar coordinates will be
\be 
ds^{2} = g_{00}(\hat{y}) dx^{0} dx^{0} + \exp(-4ak^{0}) \left[ g_{rr} (\hat{y})  dx^{r} dx^{r} + g_{\theta \theta} (\hat{y})  dx^{\theta} dx^{\theta} + g_{\phi \phi} (\hat{y})  dx^{\phi} dx^{\phi} \right].
\ee
Since, $g_{\mu \nu}(\hat{y})$ is obtained by replacing the commutative coordinates of corresponding coordinates with $\hat{y}_i$, we write down the deformed \sch \, metric as
\be
\ba 
ds^{2} =   -\left( 1-\frac{2GM}{r}\right) dt^{2} + \frac{1}{1-\frac{2GM}{r}} \exp(-4ak^{0}) \, dr^{2} + r^{2} \exp(-4ak^{0}) \, d\Omega^{2}. 
\ea
\ee
A change of coordinate given by
\be 
r \rightarrow \tilde{r} = e^{-2ak^{0}}r \label{rordash}
\ee
in the above equation will lead us to the \kd \, \sch \, metric, identical in form as in the commutative spacetime, i.e.,
\be 
\ba 
ds^{2} &=   -\left( 1-\frac{r_s}{\tilde{r}}\right) dt^{2}  + \frac{1}{1-\frac{r_s}{\tilde{r}}} \, d\tilde{r}^{2}  + \tilde{r}^{2}\, d\Omega^{2}.  \label{metricsc}
\ea
\ee
where we have defined, 
\be 
r_s = 2GM e^{-2ak^{0}}. \label{rs}
\ee 
The above expression for the \sch \, radius has a modification arising from the $\kappa$-deformation of spacetime. Because of this $a$-dependent factor, we have an exponential dependence on $k^{0}$ and thus, the effect of deformation on $r_s$ will be more prominent at smaller values of $k^{0}$'s.  We get the familiar expression for $r_s$, in the limit $a \rightarrow 0$. Note that the deformation of the $\theta, \phi$ dependent part of the deformed metric in eqn.\eqref{metricsc} is only through the overall multiplicative factor $\tilde{r}^{2}$, showing that the deformed metric is spherically symmetric as in the commutative spacetime.

We will now, for convenience, restrict to 1+1 dimensional spacetime. It has to be emphasized that all the necessary physical information, for our purpose, is retained in the two-dimensional case. An immediate advantage of working in 1+1 dimensions, is that, we could use the fact that 1+1 dimensional deformed \sch \, metric obtained is conformally flat, which simplifies our calculation. By conformally flat, we mean a metric which can be written in the form
\be 
d\tilde{s}^{2} = \Theta^{2} ds^{2} =\Theta^{2} \left( dx^{2}- dt^{2} \right) = \Theta^{2} du dv ,  \label{confflat}
\ee
where $\Theta$ is the conformal factor and $u,v$ denote the light-cone coordinates.

Before proceeding further we will now briefly outline two coordinate systems that will be of use. We write down the \kd \, Schwarzschild metric in 1+1 dimension as
\be 
ds^{2} = -\left( 1-\frac{r_s}{\tilde{r}}\right) dt^{2}  + \frac{1}{1-\frac{r_s}{\tilde{r}}} \, d\tilde{r}^{2}.  \label{metric1}
\ee
Note that the effect of $\kappa$-deformation is appearing through $\tilde{r}$ and $r_s$ (see eqns. \eqref{rordash} and \eqref{rs}). The corresponding surface gravity (surface gravity is the effective strength of gravitation at the surface of a gravitating body, see \cite{townsend} for details) is given by,
\be 
\kappa_s = \frac{1}{2 r_s} = \frac{1}{4GM e^{-2ak^{0}}}. \label{surfacegravity}
\ee
Next, we generalize the tortoise coordinates $r^{*}(r) = r - r_s + r_s ln\left(\frac{r}{r_s}-1\right)$, which is defined for $r>r_s$ to \kd \, spacetime as 
\be 
\tilde{r}^{*}(\tilde{r}) = \tilde{r} - r_s + r_s ln\left(\frac{\tilde{r}}{r_s}-1\right). \label{tortoise}
\ee
Notice that the $r_s$ appearing in the above equation is the \kd \, \sch \, radius  given in eqn.\eqref{rs} and $\tilde{r}$ is the deformed radial coordinate given in eqn. \eqref{rordash}. Changing into the \kd \, tortoise light cone coordinates,
\be 
U = t - \tilde{r}^{*}, \quad V = t+ \tilde{r}^{*},
\ee
the \kd \, metric in eq. \eqref{metric1} can be re-expressed as
\be 
ds^{2} = \left( 1-\frac{r_s}{\tilde{r}(U,V)} \right) -dU dV.
\ee
The above coordinate system (eqn. \eqref{tortoise}) is singular at $r=r_s$. We can remove this coordinate singularity by a change of variables and analytically continuing the coordinate system to cover whole of spacetime, i.e., by using variables
\be 
u = - 2r_s \exp\left( -\frac{U}{2\tilde{r}_s} \right), \quad v = 2r_s \exp\left( \frac{V}{2r_s} \right)
\ee
which are the \kd \, light cone Kruskal-Szekers coordinates. Note that the effect of deformation enter through $a$-dependence of $r_s, U$ and $V$.

In this \kd \, light cone Kruskal-Szekers coordinates, the metric takes the shape
\be 
ds^{2} = \frac{r_s}{\tilde{r}(u,v)} \exp\left( 1-\frac{\tilde{r}(u,v)}{r_s} \right) du dv. \label{metrickruskal}
\ee
Note that $r_s$ appearing here is the \sch \, radius given in eqn.\eqref{rs}. In the limit $a \rightarrow 0$, we recover the commutative metric.

We now briefly sketch the derivation of Hawking radiation in commutative space-time.  It is a known fact that vacuum is not uniquely specified in a curved spacetime. Our choice of coordinates dictate the behaviour of vacuum. In order to study the Hawking radiation, the scalar field under consideration is expanded in two different basis, namely in the Kruskal-Szekers coordinates and tortoise coordinates. Boguliobov transformation is then used to connect the creation (annihilation) operators in the two basis. Using the normalisation condition for the Boguliobov transformation, one calculates the average number density of particles in the Kruskal vacuum. The number density obtained has the form of thermal distribution of particles and this identification is then used to obtain the corresponding formula for the temperature of the Hawking radiation. Coming back to the \kd \, case, we follow the same algorithm, except that the creation (annihilation) operators obey a different algebra owing to deformation of the spacetime. This, in addition to the fact we have a modified expression for surface gravity gives us the modified Hawking temperature in \kdst .

The action of scalar theory in 2 dimensional \kdst \, can be formally written as
\be 
S = -\int d^{2}x \sqrt{-\hat{g}} \frac{1}{2} \Phi \left( \hat{g}_{\mu \nu} \tilde{D}^{\mu} \tilde{D}^{\nu} \right) \Phi.
\ee
Where $\tilde{D}$ is the covariant derivative of the deformed spacetime under consideration.

Since the metric in eqn.\eqref{metrickruskal} has the form of conformally flat metric given in eqn.\eqref{confflat}, we could replace $\hat{g}_{\mu \nu}$ with the flat metric $\eta_{\mu \nu}$ and hence the action becomes
\be
S = -\int d^{2}x \frac{1}{2} \Phi \left( \eta_{\mu \nu} D^{\mu} D^{\nu} \right) \Phi.
\ee
Euler-Lagrange equation following from the above equation, gives us the massless Klein-Gordon equation,
\be 
\left( \eta_{\mu \nu} D^{\nu} D^{\mu} \right) \Phi = 0.
\ee 
In the above equation, we have introduced a generalised derivative called Dirac derivative, denoted $D_\mu$, which transform like a vector under undeformed Poincare transformation \cite{mel}. The explicit form of the Dirac derivative is given by
\be
D_i = \partial_i \frac{e^{-A}}{\varphi}, \,\, D_0 = \partial_0 \frac{\sinh A}{A} - ia \nabla^{2} \frac{e^{-A}}{2\varphi^{2}},
\ee
where $A = -i \partial_0$.
The Casimir of the Poincare algebra has the form
\be
D_\mu D^{\mu} = \Box(1- \frac{a^{2}}{4}\Box)
\ee
with the operator $\Box$ given by
\be 
\Box = \nabla^{2} \frac{e^{-A}}{\varphi^{2}} + \frac{2\partial_0^{2}}{A^{2}}(1-\cosh A)
\ee
With identification $P_\mu = i D_\mu$, the deformed dispersion relation is given by
\be
\ba
& \frac{4}{a^{2}} \sinh^{2}(\frac{ap_0}{2}) - p_i^{2} \frac{e^{-ap_0}}{\varphi^{2}(ap_0)} + \frac{a^{2}}{4} \left[ \frac{4}{a^{2}} \sinh^{2}(\frac{ap_0}{2}) -  p_i^{2} \frac{e^{-ap_0}}{\varphi^{2}(ap_0)} \right]^{2} = 0  \label{dispkappa},
\ea
\ee
where $p_0$ and $p_i$ are energy and momentum in the commutative spacetime.

Thus the generalised Klein-Gordon equation in \ks \, is 
\be 
\left( \Box \left( 1+\frac{a^{2}}{4}\Box \right) -m^{2} \right) \Phi(x) =0. \label{kappakg}
\ee
We can decompose the field $\phi$ satisfying Klein-Gordon equation into positive and negative frequency solution, in the Kruskal-Szekeres coordinates as
\be 
\Phi(x) = \int \frac{d\omega}{2|p|} \left[ A(\omega_p, p) e^{i\omega U} + A^{\dagger}(\omega_p, p) e^{-i\omega U} \right] + \int \frac{d\omega}{2|p|} \left[ A(\omega_p, p) e^{i\omega V} + A^{\dagger}(\omega_p, p) e^{-i\omega V} \right], \label{fmod}
\ee
Here $p^{0} :=\omega_p$ is obtained from \eqref{dispkappa} with $\varphi = e^{iap^{0}}$. Note that the vector momentum $\vec{p}$ in one-dimensional space is just a number. In the case of 3+1 spacetime, we would have a vector with three components. Thus we use just $p$ and not $\vec{p}$ as in the 3+1 dimensional case as in eqn.\eqref{fmod} and equations below. We define the vacuum state, called Kruskal vacuum state, corresponding to Kruskal-Szekers coordinate as
\be 
A_\omega \left | 0_K \right > := A(\omega_k, p) |0_K> =0.
\ee
The suffix $K$ in $ | 0_K >$ is to denote that the vacuum is associated with the Kruskal-Szekeres coordinate. It is known that if we interchange two \kd \, bosonic fields, we end up with an additional factor as compared with the commutative case \cite{twist, twist2}. This would lead to a deformed commutation relation between annihilation (creation) operators. For two bosonic fields, interchange of fields in the products results in 
\be 
\phi(x) \otimes \phi(y) - e^{-(A \otimes N - N \otimes A)} \phi(y) \otimes \phi(x) = 0 \label{fflip}
\ee
where $A_x = -ia \partial_0^{x}, N_x = x_i \partial_i^{x} $ (see \cite{twist} for details). This leads to the deformed commutation relation
\bea
A(p_0,p) A(q_0,q) - e^{-a(q_0 \partial_{p} p - \partial_{q} q p_0)} A(q_0,q) A(p_0,p) =0 \label{aa}\\
A(p_0,p) A^{\dagger} (q_0,q) - e^{+a(q_0 \partial_{p} p -\partial_{q} q p_0)} A^{\dagger}(q_0,q) A(p_0,p) = \delta(p-q) \label{aad} \\
A^{\dagger}(p_0,p) A^{\dagger} (q_0,q) - e^{-a(q_0 \partial_{p} p - \partial_{q} q p_0)}  A^{\dagger}(q_0,q) A^{\dagger}(p_0,p) =  0 \label{adad}\\
A^{\dagger}(p_0,p)A(q_0,q) - e^{-a( -q_0 \partial_{p} p +  \partial_{q}  q p_0)}  A(q_0,q)A^{\dagger}(p_0,p) = -\delta(p-q)  \label{ada}
\eea
It is to be noted the combination $\partial_{q}  q= q\partial_{q}+ 1$. (The factor $1(=\frac{\partial q}{\partial q})$ appear because we are working in a spacetime with one spatial coordinate. Contrast this with the factor 3 appearing in eqn.(87-90) in \cite{twist2}).
The expressions given by eqns.\eqref{aa}-\eqref{ada} are \kd \, analog of commutation relations of creation and annihilation operators in the commutative spacetime. One can define a deformed product called ``o-product'' to re-express eqn. (\eqref{aad}) in a compact way as in \cite{twist}, i.e., 
\be 
A^{\dagger}(p) \circ A(q) = e^{\frac{a}{2}(p_0-q_0)} A^{\dagger}(e^{+a\frac{\omega_q}{2}}p) A(e^{-a\frac{\omega_p}{2}}q). \label{oproduct}
\ee

Consider now a mode expansion of field $\Phi(x)$ satisfying eqn. \eqref{kappakg} in tortoise coordinates. We obtain
\be
\Phi(x) = \int  \frac{d\Omega}{2|p|} \left[ B(p^{\prime}_0,p) e^{i\Omega u} + B^{\dagger}(p^{\prime}_0,p)e^{-i\Omega u}  \right] + \int  \frac{d\Omega}{2|p|} \left[ B(p^{\prime}_0,p) e^{i\Omega v} + B^{\dagger}(p^{\prime}_0,p)e^{-i\Omega v}  \right] \label{phiB}
\ee
where $p^{\prime}_0 = \Omega_p = \kappa_s \omega_p$, with the dependence on the deformation parameter, $a$ coming from the the surface gravity term $\kappa_s$(see eqn. \eqref{surfacegravity}) and from the expression for $p_0 =\omega_p$ given by eqn.\eqref{dispkappa}. Note that $\Omega_p$ is the energy in the deformed Kruskal-Szekers coordinate system and $\omega_p$ is the energy in the deformed tortoise coordinates given in eqn. \eqref{dispkappa} (See for example, chapter 8 of \cite{mukh}). Note that $B$ and $B^{\dagger}$ appearing in eqn. \eqref{phiB} also obey twisted commutation relations as $A, A^{\dagger}$. These relations can be obtained by replacing $A$ and $A^{\dagger}$ with $B$ and $B^{\dagger}$ in eqn. \eqref{aa} to \eqref{ada}, respectively. This will allow us to define modified o-product for $B$ and $B^{\dagger},$ which will have exactly the same form as in eqn.\eqref{oproduct}. We can also define the vacuum state corresponding to the tortoise coordinates, called Boulware vacuum, as
\be 
B_\Omega |0_B> := B(p^{\prime}_0,p)|0_B> =0.
\ee
Now, $B(p^{\prime}_0, p)$ can be expressed in terms of $A$ and $A^{\dagger}$ in the form
\be 
A(\omega, p) = \int d\Omega \left( \alpha_{\Omega \omega} B(\Omega, p) + \beta_{\Omega \omega} B^{\dagger}(\Omega, p) \right) \label{bog}
\ee
where $\alpha, \beta$ are known as Boguliobov coefficients. Further, the inverse Boguliobov transformation is given by
\be 
B(\Omega, p) = \int d\omega \left( \alpha_{\Omega \omega} A(\omega, p) - \beta_{\Omega \omega} A^{\dagger}(\omega, p) \right) \label{invbog}
\ee

Now demanding that $A \, (B)$ obey the eqns. (\eqref{aa}-\eqref{adad}), we will derive the relations satisfied by the coefficients, $\alpha$ and $\beta$. We start with the eqn. \eqref{aad}, and proceed to understand the properties of Boguliobov coeffecient.

Using eqn. \eqref{bog}, L.H.S. of eqn. \eqref{aad} will take the form
\be 
\ba 
& \int d\Omega_p d\Omega_q \, \left[ \alpha B(p) + \beta B^{\dagger}(p)\right] \left[ \alpha^{*} B^{\dagger}(q)+ \beta^{*}B(q)\right] - \\ & e^{+a(q_0 \partial_{p} p -\partial_{q} q p_0)} \left[\alpha^{*} B^{\dagger}(q)+ \beta^{*}B(q)\right] \left[\alpha B(p) + \beta B^{\dagger}(p)\right]
\ea
\ee
which is written as
\be 
\ba 
{} & \int d\Omega_p d\Omega_q \, \alpha \alpha^{*} B(p) B^{\dagger}(q) - e^{+a(q_0 \partial_{p} p -\partial_{q} q p_0)} \alpha \alpha^{*} B^{\dagger}(q) B(p) \\ & + \alpha B(p)\beta^{*}B(q) - e^{+a(q_0 \partial_{p} p -\partial_{q} q p_0)} \alpha^{*} B^{\dagger}(q)\beta B^{\dagger}(p)) \\ & + \beta B^{\dagger}(p)\alpha^{*} B^{\dagger}(q) - e^{+a(q_0 \partial_{p} p -\partial_{q} q p_0)} \beta^{*}B(q)\alpha B(p) \\& + \beta B^{\dagger}(p)\beta^{*}B(q) - e^{+a(q_0 \partial_{p} p -\partial_{q} q p_0)}\beta^{*}B(q)\beta B^{\dagger}(p). \label{eq53}
\ea
\ee
With the help of eqns. \eqref{aa}-\eqref{adad}, we get from eqn.\eqref{eq53}
\be 
\ba 
{} & \int d\Omega_p d\Omega_q \, \alpha \alpha^{*} \delta(\Omega_p - \Omega_q) - \beta \beta^{*}  \delta(\Omega_p - \Omega_q) \\ & + \alpha \beta^{*} \left( B(p)B(q) - e^{+a(q_0 \partial_{p} p -\partial_{q} q p_0)} B(q) B(p) \right) \\ & \beta \alpha^{*} \left( B^{\dagger}(p) B^{\dagger}(q) - e^{+a(q_0 \partial_{p} p -\partial_{q} q p_0)}  B^{\dagger}(q) B^{\dagger}(p)) \right)\\  & = \delta(\omega_p - \omega_q) .
\ea
\ee
or
\be 
\ba 
{} & \int  d\Omega_p \, \left( \alpha_{\Omega_p \omega_p} \alpha^{*}_{\Omega_p \omega_q} - \beta_{\Omega_p \omega_p} \beta^{*}_{\Omega_p \omega_q} \right) + \int d\Omega_p d\Omega_q \\ & \alpha \beta^{*} \left( B(p)B(q) - e^{+a(q_0 \partial_{p} p -\partial_{q} q p_0)} B(q) B(p) \right) + \\ & \beta \alpha^{*} \left( B^{\dagger}(p) B^{\dagger}(q) - e^{+a(q_0 \partial_{p} p -\partial_{q} q p_0)}  B^{\dagger}(q) B^{\dagger}(p)) \right)\\  & = \delta(\omega_p - \omega_q) . \label{bogrel}
\ea
\ee
Using eqn. \eqref{invbog} in \eqref{phiB}, we obtain an expression for $\Phi(x)$ in terms of $A(p_0, p)$ and Boguliobov coefficients. But this expression should be equal to eqn.\eqref{fmod}. Equating the two expressions and carrying out straightforward algebra, gives us
\bea
\alpha_{\Omega \omega} = \frac{1}{2\pi \kappa_s} \sqrt{\frac{\Omega}{\omega}} e^{\frac{\pi \Omega}{2\kappa_s}} \exp\left( \frac{i\Omega}{\kappa_s} ln \frac{\omega}{\kappa_s}\right) \Gamma(-\frac{i\Omega}{\kappa_s}), \\
\beta_{\Omega \omega} = - \frac{1}{2\pi \kappa_s} \sqrt{\frac{\Omega}{\omega}} e^{-\frac{\pi \Omega}{2\kappa_s}} \exp\left( \frac{i\Omega}{\kappa_s} ln \frac{\omega}{\kappa_s}\right) \Gamma(-\frac{i\Omega}{\kappa_s})\label{bogrel23}
\eea 
Using the ``o-product'' defined for $B$, similar to the one defined in eqn. \eqref{oproduct}, expression for the number operator in the tortoise coordinates takes the form
\bea 
N_T(p) &=& B^{\dagger}(p) \circ B(p) = B^{\dagger}(e^{+a\frac{\Omega}{2}}p) B(e^{-a\frac{\Omega}{2}}p).
\eea
Here, the subscript $T$ is to emphasis that the number operator is associated with the tortoise coordinates. From Eqn. \eqref{invbog}, we find the expectation value of this number operator in Kruskal vacuum to be 
\be 
\ba 
<N_T(p)> & = <0_K| \int d\omega_p d\omega_q \beta^{*}_{\Omega \omega}\beta_{\Omega \omega^{\prime}}  A(e^{+a\frac{\omega_p}{2}}p) A^{\dagger}(e^{-a\frac{\omega_p}{2}}p)|0_K>
\ea
\ee
This immediately gives us
\be 
<N_T(p)> = \int d\omega |\beta_{\Omega \omega}|^{2}. \label{num1}
\ee
The condition \eqref{bogrel}, for the case $\Omega_p = \Omega_q$ and $p=q$, and $p^{0}=q^{0}$, will be simplified to
\be 
\int d\Omega \left( \left| \alpha_{\Omega \omega} \right|^{2} - \left| \beta_{\Omega \omega} \right|^{2} \right) = \delta(0). \label{bogrel3}
\ee
Taking into account of eqn.\eqref{bogrel23}, we obtain from eqn.\eqref{bogrel3},
\be 
<N_T(p)> = \int d\omega |\beta_{\Omega \omega}|^{2} = \left[\exp\left(\frac{2\pi \Omega}{\kappa_s}\right)-1 \right]^{-1} \delta(0).
\ee
Thus, average particle density in the Kruskal vacuum is 
\be 
n(\Omega) = \frac{<N_T(p)>}{V} = \left[\exp\left(\frac{2\pi \Omega}{\kappa_s}\right)-1 \right]^{-1} .
\ee
The expression can be easily identified with Bose-Einstein distribution having temperature
\be 
T_{H} = \frac{\kappa_s}{2\pi} = \frac{1}{8\pi GM e^{-2ak^{0}}}. \label{th1}
\ee
We have thus obtained a thermal distribution for a \kd -\sch \, black hole. We thus see that a black hole in \kdst \, could radiate with effective temperature which depends inversely on its mass as in the commutative case. Notice that we have a $k^{0}$ dependence on our metric. Demanding that the $k^{0}$ satisfies the dispersion relation ${k^{0}}^{2}=\vec{k}^{2}+m^{2}$, we can interpret $k^{0}$ as the energy scale. The significance of our result is that the Hawking temperature now has a dependence on the energy scale. Since $k^{0}$ is the energy scale appearing in the metric and therefore, it must be the energy scale at which the black hole is formed. Note that eqn.\eqref{th1} suggest that a black hole formed at a higher energy scale will a have a higher $T_H$ than another black hole formed at a lower energy scale. Also note that the Hawking temperature will be higher than that in commutative case (for $a >0$). In the limit $a \rightarrow 0$, this $k^{0}$ dependence vanishes and we recover the usual result as expected. For $a >0$, we see that $T_H$ increases whereas for $a<0$, it decreases (we have taken $k^{0}>0$).

Employing the formula, 
\be 
T_H = \frac{\text{surface gravity}}{2\pi}.
\ee
(Validity of this expression in non-commutative spacetime is explained in appendix B.) Now, we infer that the Hawking temperature corresponding to the metric given in eqn. \eqref{metric1} to be 
\be 
T_H = \frac{\text{1}}{4\pi r_s} = \frac{1}{8\pi GM e^{-2ak^{0}}} \label{th}
\ee
with deformed surface gravity $\kappa_s = 1/2r_s$. This equation matches exactly with the expression (eqn. \eqref{th1}). This confirms that the result obtained is valid even for the case of 3+1 dimensions, even though our calculations leading to eqn. \eqref{th1} is carried out in 1+1 dimensions.

\section{Discussions and conclusion}
Study of structure of spacetime is attracting interest these days as it can shed light into an understanding of quantum theory of gravity. In the present paper, we have used the framework of \kdst \, to study the spacetime around a \kd -\sch \, black hole. It is well known that black hole has a temperature and it radiates thermally, when we take into account the quantum aspects of black holes. In the present paper, we have studied the radiation of \kd - \sch \, black hole. A more formal approach to understand the \kd \, would be to derive the non-commutative Einstein equation and use the solution to this \kd \, Einstein equation and analyze the \kd \, black hole. However, in this paper we have used a minimalistic approach to deal with \kd - \sch . We have considered a non-commutative generalisation of Schwarzschild black hole by demanding that we retain the commutative Schwarzschild solution as we let the $\kappa$-deformation parameter to vanish. For this, we started with a realization of \ks \, variables in terms of commutative variables. This realization was then used to construct a deformed metric using commutation relations of the phase space coordinate associated with the $\kappa$-spacetime. The main features of our metric are: (i) the \kd \, metric has additional terms depending on the deformation parameter $a$. The presence of these factors reveal the extend of deformation of the spacetime, (ii) the \kd \, metric is not symmetric under the interchange of its indices, (iii) the resulting metric has a $k^{0}$ dependence, which can be interpreted as dependence on the energy scale of the spacetime. It should be emphasized that all these features  of the metric due to the $\kappa$-deformation vanish when the deformation parameter is set to zero. Next, we have used these results to set up a \kd -\sch \, metric and thermal radiation of the deformed \sch \, black hole is calculated. This is done by studying \kd \, massless scalar field in the background of \kdst \, with Schwarzschild metric. We have shown that the 1+1 dimensional \kd \, \sch \, metric is conformally flat. This allows us to set up the scalar field theory in the \sch \, background by replacing the \kd \, \sch \, metric with the scalar theory in the \kd \, Minkowski spacetime. We then Fourier expanded the Klein-Gordon field, both in the  \kd \, Kruskal-Szekers coordinates and \kd \, tortoise coordinates. Using the fact that the creation (annihilation) operators in both coordinates are related by a Boguliobov transformation, we derived relations between creation (annihilation) operators of Kruskal-Szekers coordinates and tortoise coordinates. Bogiliobov coefficients are then obtained and this result is used to calculate the particle number density by evaluating the expectation value of the number operator in \kd \, tortoise coordinates on \kd \, Kruskal vacuum. Particle number so calculated led us to an expression for modified thermal radiation. This result is used to identify the Hawking temperature of deformed \sch \, black hole.

Using the modified Hawking temperature, we calculate the entropy of deformed \sch \, black hole (see for eg. \cite{townsend}),
\bea 
S = c^{2} \int \, \frac{dM}{T_H} = \int \, dM \frac{8 \pi  G k_B M  \exp \left(\frac{a k^{0}}{c \hbar}\right)}{hc}
= \frac{4 \pi  G k_B M^{2} \exp \left(\frac{-2a k^{0}}{c \hbar}\right)}{hc^3}. \label{S71}
\eea
From now onwards, we set the value of Boltzmann constant to one. For the case of \sch \, black hole, the area of horizon is given by 
\be 
\mathcal{A} = 4 \pi r_s^{2} = \frac{16 \pi G^{2} M^{2} \exp \left(\frac{-4a k^{0}}{c \hbar}\right)}{c^{4}}
\ee
and thus the expression in eqn.\eqref{S71} for the entropy takes the form
\be 
S =  \frac{c^{3}\mathcal{A}}{4G\hbar \exp \left(\frac{-2a k^{0}}{c \hbar}\right)}. \label{NCS}
\ee
Note that the entropy has a dependence on energy scale $k^{0}$ as well as on deformation parameter. We thus have a correction to entropy resulting from the deformation of the spacetime. Note that in the limit $a \rightarrow 0$, we recover Hawking-Bekenstein formula 
\be 
S =  \frac{c^{3}\mathcal{A}}{4G\hbar}.
\ee
From eqn. \eqref{NCS}, we note that $S$ increases as $k^{0}$ increases, for fixed values of non-commutative parameter $a$. An analysis of the specific heat is also helpful in understanding the  thermodynamical nature of black hole. The specific heat (in natural units) is given by
 \be 
 C = \frac{\partial T_H}{\partial M} = - \frac{1}{8\pi M^{2} e^{-2ak^{0}}}.
 \ee
 We see that system is thermodynamically unstable, as in the commutative case, since the temperature rises as the mass decreases. The presence of the exponential factor $e^{-2ak^{0}}$ means that \kd \, specific heat will be more negative at higher energy levels (for $a>0$), compared with that of the  commutative spacetime. We also see that the expression matches identically with the commutative spacetime result when we take the deformation parameter to zero.

We will now analyze the effect of deformation at various energy scales. Before proceeding, we will write down eqn. \eqref{th} 
\be 
T_H = \frac{hc^3}{8 \pi  G k_B M \exp \left(\frac{-2a k^{0}}{c \hbar}\right)},
\ee
where we brought back the Boltzmann's constant, $k_B$ . We recognize that the factor $e^{\frac{-2ak^{0}}{c \hbar}}$ to be crucial in determining the effect of $\kappa$-deformation of spacetime on Hawking temperature. A striking feature of our result is that, in the \kdst , the Hawking temperature has a dependence on the energy scale at which we probe the system. See the following fig.(\ref{fig:figure1}) for an illustration of the deviation of Hawking temperature due to deformation of the spacetime. From the figure, we observe that the Hawking temperature varies more from the commutative result as we move up in energy (note that we have plotted energy ($k_0$ in joules) vs Hawking temperature ($T_H$ in kelvin) in these figures).  If we take up the example of a black hole of mass of $10^{6} M_{\odot}$, we see that variation of $T_H$ with energy is more prominent when the order of deformation parameter is of larger magnitude. Also, notice that, for  deformation parameter of the order of magnitude of $10^{-40}$, the difference between deformed Hawking temperature and the Hawking temperature of commutative spacetime is negligible for the current accessible energy scales. This means that at the energy scale considered, effects of deformation is more prominent only if $a > 10^{-40}$. 
 \begin{figure}[!htb]
\centering
\begin{minipage}{.5\textwidth}
\centering
\includegraphics[scale=.5]{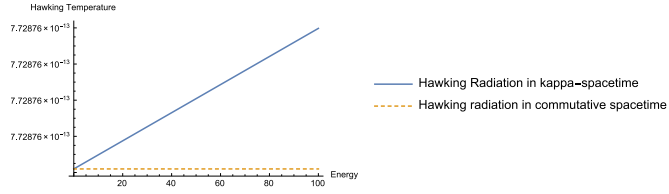}
\caption{Variation of Hawking temperature, for a black hole of mass $10^{6} M_{\odot}$ with energy, for the case $a\sim 10^{-34}m$}
\label{fig:figure1}
\end{minipage}
\end{figure}
 
While studying the deformation of metric in the \kdst , we found that the metric, in general, will be asymmetric in cross terms of space and time coordinates (see $\hat{g}_{oi}$ and $\hat{g}_{io}$ in eqn.\eqref{g0i} and\eqref{gi0}). For a general deformed metric, we see that the metric will have the form of usual commutative spacetime metric when we take the limit $k^{0}\rightarrow \infty$ and $a \rightarrow 0$, by keeping the product $k^{0}a$ to be finite. For the limit $k^{0}\rightarrow \infty$ with $a\neq 0$, the only non-vanishing component of metric is $\hat{g}_{00}$. If we let $k^{0} \rightarrow 0$ by keeping $a\neq 0$, then the only deformation in the metric arise from the functional form of realization of $\hat{y}_\mu$. In this limit, we see from eqn. \eqref{th1} that $T_H$ is same as that of the commutative case irrespective of the value of $a$. But one may still have the effect of deformation on $T_H$ through deformed Newton's constant $G$ (see \cite{hari1, renG}). Also, one may get a dependence through a deformed $m$ appearing in eqn. \eqref{th1} (see \cite{zns,zns1,zns2,akk}). Here we focused on the effect of deformed \sch \, metric in \ks \, on Hawking radiation. Since, the deformed \sch \, metric does not have any off-diagonal terms, it is completely symmetric under the interchange of indices. The \kd \, Schwarzschild metric, has a modification compared with the commutative case, arising from the presence of factors with a dependence in the deformation parameter $a$. This deformation enter the metric through the dependence of $\tilde{r}$ and $r_s$ (see eqn. \eqref{rordash} and \eqref{rs}). Note that the dependence of $\tilde{r}$ and $r_s$ on $a$ is through the combination $ak^{0}$. Note that the deformed metric do not have any $a$ dependence that is independent of $k^{0}$. This is the reason of the dependence of $T_H$ on $ak^{0}$. As stated, dependence on $a$, independent of $k^{0}$, can appear through a deformed Newton's constant \cite{hari1,renG} and deformed mass parameter \cite{akk}. Also, we have found that due to presence of deformation of metric, the surface gravity for a \kd -Schwarzchild black hole gets modified. Investigation of thermal radiation in the presence of a Schwarzschild-like metric have shown modification of temperature of Hawking radiation. Our calculation gives a correction to entropy of \kd \, \sch \, black hole so that the resultant entropy is higher than the commutative result, for a given energy scale (with $a>0$). For the case, where the deformation of the spacetime vanishes, correction to the entropy vanishes and the entropy reduces to familiar expression. Thus, it seem that Hawking temperature of the black holes formed, when the energy density of universe is higher, were higher than the black holes formed latter. As $a\rightarrow 0$, $T_H$ of all black holes become independent of the energy scale. This may have implications for the primordial black holes. It has been argued that a distinction between Hawking effect and Unruh effect is possible in a consistent way (see \cite{barcelo} and refereces therin). It will be of interest to see how this effect is affected by non-commutativity of spacetime. 

\bigskip

\noindent{\bf Acknowledgements.}  ZNS would like to acknowledge the support for this work from CSIR, India through the SRF scheme. EH thanks SERB, Govt. of India, for support through EMR/2015/000622.
\begin{appendices}
\section{Deformed metric computation}
In this appendix, we derive a realization for auxiliary variable $\hat{y}^{\mu}$ introduced in Section II. We start with conditions satisfied by $\hat{y}^{\mu}$ given by eqns. \eqref{ycomm},\eqref{cond2}, to express the non-commutative variable $\hat{y}^{\mu}$ in terms of commutative variables. Thus we start with 
 \bea
 0= [ \hd{y}{0}, \hd{x}{0} ] &=& [x_\alpha \phi^{\alpha}_0, x_\beta  \varphi_{0}^{\beta}] = x_\beta [x_\alpha, \varphi_{0}^{\beta}] \phi^{\alpha}_0 + x_\alpha [\phi^{\alpha}_0 , x_\beta]  \varphi_{0}^{\beta}, \\
&=&  x_\beta \varphi_{0, \alpha}^{\beta} \phi^{\alpha}_0 - x_\alpha \phi^{\alpha}_{0, \beta}  \varphi_{0}^{\beta}, =  x_\beta \varphi_{0, 0}^{\beta} \phi^{0}_0 - x_\alpha \phi^{\alpha}_{0, 0}  \varphi_{0}^{0} \eea
 where we have used the fact that $\varphi_0^{i}=0$ (eqn. \eqref{real}). Also, note that the commutators between $\phi^{\alpha}_\beta$ and $\varphi^{\alpha}_\beta$ are identically zero as both of the realizations are function of $p_\mu$. We thus have 
\be 
 \phi^{\alpha}_{0, 0}  =0. \label{c1}
 \ee
Starting with $0= [ \hd{y}{0}, \hd{x}{i} ] $, we get
 \bea
  0= [ \hd{y}{0}, \hd{x}{i} ] &=& [x_\alpha \phi^{\alpha}_0, x_\alpha  \varphi_{i}^{\alpha}] = x_\beta [x_\alpha, \varphi_{i}^{\beta}] \phi^{\alpha}_0 + x_\alpha [\phi^{\alpha}_0 , x_\beta]  \varphi_{i}^{\beta} \\
&=&  x_j \varphi_{i, 0}^{j} \phi^{0}_0 - x_0 \phi^{0}_{0, j}  \varphi_{i}^{j} - x_k \phi^{k}_{0, j}  \varphi_{i}^{j}
 \eea
 This implies that
 \be 
  \phi^{0}_{0, j}  =0 \, \, \text{and} \, \,  \phi^{k}_{0, j} + a \delta^{k}_j \phi^{0}_0  =0. \label{c2}
 \ee
 Next, we start with $0= [ \hd{y}{i}, \hd{x}{0} ]$,
 \bea
 0= [ \hd{y}{i}, \hd{x}{0} ] &=& [x_\alpha \phi^{\alpha}_i, x_\alpha  \varphi_{0}^{\alpha}] 
 = x_\beta [x_\alpha, \varphi_{0}^{\beta}] \phi^{\alpha}_i + x_\alpha [\phi^{\alpha}_i , x_\beta]  \varphi_{0}^{\beta} \\
&=&  x_\beta \varphi_{0, \alpha}^{\beta} \phi^{\alpha}_i - x_\alpha \phi^{\alpha}_{i, \beta}  \varphi_{0}^{\beta} =  
\cancel{x_0 \varphi_{0, 0}^{0} \phi^{0}_i} - \cancel{x_\alpha \phi^{\alpha}_{i, j}  \varphi_{0}^{j}} 
-x_\alpha \phi^{\alpha}_{i, 0}  \varphi_{0}^{0},
\eea
and obtain the condition on $\phi^{\alpha}_{i, 0}$ as
 \be 
 \phi^{\alpha}_{i, 0} =0. \label{c3}
 \ee
 The commutation relation $[ \hat{y}_i, \hat{x}_j ] $ gives,
 \bea
 0= [ \hd{y}{i}, \hd{x}{j} ] &=& [x_\alpha \phi^{\alpha}_i, x_\alpha  \varphi_{j}^{\alpha}] = x_\beta [x_\alpha, \varphi_{j}^{\beta}] \phi^{\alpha}_i + x_\alpha [\phi^{\alpha}_i , x_\beta]  \varphi_{j}^{\beta} \\
&=&  x_\beta \varphi_{j, \alpha}^{\beta} \phi^{\alpha}_i - x_\alpha \phi^{\alpha}_{i, \beta}  \varphi_{j}^{\beta} =  x_k \varphi_{j, 0}^{k} \phi^{0}_i - x_l \phi^{l}_{i, k}  \varphi_{j}^{k} - x_0 \phi^{0}_{i, k}  \varphi_{j}^{k}
\eea
This gives us the condition
 \be 
\phi^{0}_{i, k}  =0 \, \, \text{and} \, \,  \phi^{l}_{i, j} + a \delta^{l}_j \phi^{0}_i  =0. \label{c4}
 \ee
 Now, using eqn. \eqref{def1}, we get
 \be
 0 =[\hd{y}{0}, \hd{y}{0}] = x_\beta [x_\alpha, \phi_{0}^{\beta}] \phi^{\alpha}_0 
 + x_\alpha [\phi^{\alpha}_0 , x_\beta]  \phi_{0}^{\beta} 
=  x_\beta \phi_{0, \alpha}^{\beta} \phi^{\alpha}_0 - x_\alpha \phi^{\alpha}_{0, \beta}  \phi_{0}^{\beta} =0. 
 \ee
 Using the realization for $\hat{y}_0, \hat{y}_i$ in $ \hd{y}{i} = [\hd{y}{0}, \hd{y}{i}] $ gives
  \be
a x_\alpha \phi^{\alpha}_i  = x_\beta [x_\alpha, \phi_{i}^{\beta}] \phi^{\alpha}_0 
+ x_\alpha [\phi^{\alpha}_0 , x_\beta]  \phi_{i}^{\beta}
=  x_\beta \phi_{i, \alpha}^{\beta} \phi^{\alpha}_0 - x_\alpha \phi^{\alpha}_{0, \beta}  \phi_{i}^{\beta}, 
\ee
After simplification, we get the following condition from the above equation,
\be 
 a (x_0 \phi^{0}_i + x_j \phi^{j}_i )
= x_j \phi_{i, k}^{j} \phi^{k}_0 - x_j \phi^{j}_{0, k}  \phi_{i}^{k}. \label{d1}
 \ee
 This sets $ \phi^{0}_i =0.$ Let us now gather together the results to obtain a realization for $\hat{y}$.
 \bea
 \phi^{\alpha}_{0,0} =0 \, \text{and} \,  \phi^{0}_{0,j} =0 \Rightarrow  \phi^{0}_{0} = const. = k_1 \nonumber \\
  \phi^{\alpha}_{i,0} =0 \, \text{and} \, \phi^{0}_{i,j} =0 \Rightarrow \phi^{0}_{i} = const =0.
 \eea
 Using eqn. \eqref{c4} and \eqref{c3}, we can write $\phi^{i}_{j} = const. = k_2 \delta^{i}_j $. Finally, condition, \eqref{c2} give us,
 \be 
 \phi^{k}_{0} = - a k_3 \delta^{k}_j \phi^{0}_0 p^{j}.
 \ee
 where $k_3$ is an integration constant. We can now write,
 \bea
 \hd{y}{0} &=& x_0 \phi^{0}_0 + x_i \phi^{i}_0 = k_1 x_0 -a k_1 k_3 x_k \delta^{k}_j  p^{j}, \\
  \hd{y}{i} &=& x_0 \phi^{0}_i + x_j \phi^{j}_i = k_2 x_j \delta^{j}_i .
 \eea
 We will now fix the constants by substituting in eqn. \eqref{def1},
 \bea 
 [ \hd{y}{0},  \hd{y}{i}] &=&  [k_1 x_0 - x_i k_3 k_1 a \delta^{k}_j p^{j}, k_2 x_j \delta^{j}_i ] \\
 &=& -k_1 k_2 k_3  [ x_k \delta^{k}_j p^{j}, x_l \delta^{l}_i ] = -a k_1 k_2 k_3  x_k  \delta^{k}_j \delta^{l}_i (-i) \delta^{j}_l = ia  k_1 k_2 k_3 x_i.
 \eea
 Hence, we get the condition that $k_1 k_3 =1$ and $k_2$ is arbitrary. We choose $k_2 =1$ and $k_1, k_3=1$ to have the correct commutative limit. With the commutation relation, $[\hat{y}_i, \hat{y}_j]=0$, we can immediately verify that the realization we obtained is consistent. We thus write the expression for $\hat{y}$ as
 \be
  \hd{y}{0} = x_0 - a x_j p^{j}, ~~
  \hd{y}{i} =  x_i.
 \ee
 \section{Hawking temperature using imaginary time method}
We start with a spherically symmetric metric in the deformed spacetime, 
\be 
ds^2=-f(r,a)\,dt^2+f^{-1}(r,a)\,dr^2+r^2 g(a)\,d\Omega^2.
\ee
with $f(r,a)$ is a function with dependence on the deformation parameter $a$ and invariant under 3-dimensional rotations. $g(a)$ is an arbitrary function of deformation parameter and independent of $r, \theta, \phi$.

Our idea is to expand the metric around the point of singularity and for that let (see \cite{euchawk} for more details)
$f(r_*,a)=0$ and now expand $f(r,a)$ near $r=r_*$. With $r= r_* + \upsilon$ ($\upsilon << 1$).
\be 
f(r_* +\upsilon)\simeq \upsilon f^\prime(r_*)
\ee
and our metric will then take the form (using $\upsilon = \epsilon^{2}$ for convenience),
\be 
ds^2=-\epsilon^2f^\prime(r_*,a )\,dt^2+4\epsilon^2\left[\epsilon^2f^\prime(r_*,a)\right]^{-1}\,d\epsilon^2+(r_* +\epsilon^2)^2\, g(a) \, d\Omega^2
\ee
With the substitution, $t = i \tau$,
\be 
\ba ds^2&=\epsilon^2f^\prime(r_*,a)\,d\tau^2+4\epsilon^2\left[\epsilon^2f^\prime(r_*,a)\right]^{-1}\,dr^2+(r_*+\epsilon^2)^2\,d\Omega^2\\&=\frac{4}{f^\prime(r_*,a)}\left[d\epsilon^2+\left(\frac{f^\prime(r_*,a)}{2}\right)^2\epsilon^2d\tau^2\right]+(r_*+\epsilon^2)^2\, g(a)\, d\Omega^2 \ea.
\ee
The coordinate $\tau$ is not restricted in range and hence the circumference of a circle of radius $\rho$ in this metric will not be $2\pi \rho$. In other words, the resulting spacetime is a cone rather than flat. The condition to avoid a conical singularity will be
\be 
\frac{\left|f^\prime(r_*,a)\right|}{2}\int_0^\beta\int_0^{r_*}\epsilon\,d\epsilon\,d\tau :=\pi{r}_*^2\,\Longrightarrow\, \beta^{\prime} =\frac{4\pi}{\left|f^\prime(r_*,a)\right|}. \label{betaprime}
\ee
where $\beta^{\prime}$ is the periodicity of $\tau$. Another way to look at this condition as follows. Let us define a quantity, $\Theta= \frac{f^\prime(r_*,a) \tau}{2}$. Then, the metric expressed in terms of $\Theta$ will be that of a sphere of radius $\epsilon$. Hence, we can demand periodicity for our angle variable, i.e., 
\be 
\Theta \approx \Theta + 2\pi \Rightarrow \tau \approx \tau + 
\frac{4\pi}{f^\prime(r_*,a)}.
\ee
 Following the connection between path integral formulation in Euclidean time and statistical mechanics, we can write the partition function of a system at temperature, $\beta$ as
\be 
Z(\beta) = Tr e^{-\beta H} = \int_{q(0)=q(\beta)} \, \mathcal{D}q e^{-S_E(q)}.
\ee
This immediately tells us that the periodicity $\beta^{\prime}$ (given in eqn. \eqref{betaprime}) can be identified with inverse temperature. It should be emphasized that we are identifying the periodicity, which depends on the deformation parameter $a$, with the temperature $\beta$ in the above expression and thus the resulting temperature will have a deformation dependence.

In our case, $f(r,a)= 1-\frac{2GMe^{-2ap^{0}}}{r}$ and we get, $f^\prime(r_*) = \frac{1}{2GMe^{-2ap^{0}}}$, where $r_* = 2GMe^{-2ap^{0}}$. Recognising, $T = \frac{1}{\beta}$ we arrive at the expression for temperature given in eqn. \eqref{th1}.
\end{appendices}

\end{document}